\documentclass{article}
\usepackage{spconf,amsmath,epsfig}
\usepackage{graphicx}
\usepackage{epstopdf}
\usepackage{multirow}

\usepackage{enumitem}
\usepackage{url}
\usepackage{hyperref}

\makeatletter
\def\bstctlcite{\@ifnextchar[{\@bstctlcite}{\@bstctl
cite[@auxout]}}
\def\bstctlcite[#1]#2{\@bsphack
\@for\@citeb:=#2\do{%
\edef\@citeb{\expandafter\@firstofone\@citeb}%
\if@filesw\immediate\write\csname #1\endcsname{\s
tring\citation{\@citeb}}\fi}%
\@esphack}
\makeatother

\begin{document}\sloppy

\onecolumn 

\begin{description}[labelindent=1cm,leftmargin=3cm,style=multiline]

\item[\textbf{Citation}]{D. Temel and G. AlRegib, "BLeSS: Bio-inspired low-level spatiochromatic similarity assisted image quality assessment," 2016 IEEE International Conference on Multimedia and Expo (ICME), Seattle, WA, 2016, pp. 1-6.
} \\

\item[\textbf{DOI}]{\url{https://doi.org/10.1109/ICME.2016.7552874}} \\

\item[\textbf{Review}]{Date added to IEEE Xplore: 29 August 2016} \\

\item[\textbf{Code/Poster}]{\url{https://ghassanalregib.com/publications/}} \\

\item[\textbf{Bib}] {
@INPROCEEDINGS\{Temel2016\_BLeSS,\\ 
author=\{D. Temel and G. AlRegib\},\\ 
booktitle=\{2016 IEEE International Conference on Multimedia and Expo (ICME)\}, \\
title=\{BLeSS: Bio-inspired low-level spatiochromatic similarity assisted image quality assessment\},\\ 
year=\{2016\},\\
pages=\{1-6\}, \\
doi=\{10.1109/ICME.2016.7552874\},\\ 
ISSN=\{1945-788X\}, \\
month=\{July\},\}\\
} \\

\item[\textbf{Copyright}]{\textcopyright 2016 IEEE. Personal use of this material is permitted. Permission from IEEE must be obtained for all other uses, in any current or future media, including reprinting/republishing this material for advertising or promotional purposes,
creating new collective works, for resale or redistribution to servers or lists, or reuse of any copyrighted component
of this work in other works. }

\item[\textbf{Contact}]{\href{mailto:alregib@gatech.edu}{alregib@gatech.edu}~~~~~~~\url{https://ghassanalregib.com/}\\ \href{mailto:dcantemel@gmail.com}{dcantemel@gmail.com}~~~~~~~\url{http://cantemel.com/}}
\end{description} 

\thispagestyle{empty}
\newpage
\clearpage

\twocolumn

\def\x{{\mathbf x}}
\def\L{{\cal L}}

\title{BL\MakeLowercase{e}SS: Bio-inspired low-level spatiochromatic similarity  assisted image quality assessment}
%
\name{Dogancan Temel and Ghassan AlRegib}
\address{Center for Signal and Information Processing (CSIP)\\
School of Electrical and Computer Engineering\\
Georgia Institute of Technology, Atlanta, GA, 30332-0250 USA\\
\{cantemel,alregib\}@gatech.edu}

\maketitle

\begin{abstract}
This paper proposes a biologically-inspired low-level spatiochromatic-model-based similarity method (\textbf{BLeSS}) to assist full-reference image-quality estimators that originally oversimplify color perception processes. More specifically, the spatiochromatic model is  based on spatial frequency, spatial orientation, and surround contrast effects. The assistant similarity method is used to complement image-quality estimators based on phase congruency, gradient magnitude, and spectral residual. The effectiveness of \textbf{BLeSS} is validated using FSIM, FSIMc and SR-SIM methods on LIVE, Multiply Distorted LIVE, and TID 2013 databases. In terms of Spearman correlation, \textbf{BLeSS} enhances the performance of all quality estimators in color-based degradations and the enhancement is at $100\%$ for both feature- and spectral residual-based similarity methods. Moreover, \texttt{BleSS} significantly enhances the performance of SR-SIM and FSIM in the full TID 2013 database.
\end{abstract}
\section{Introduction}
\label{sec:intro}
\vspace{-2.0mm}

In the 1990s, people used to spend significant amount of time before capturing photos. Because, they only had limited ($24$, $36$) exposures per roll. Nowadays, users photograph a scene with a single touch using smart devices, which capture multiple shots to provide a variety of options. Therefore, the bottleneck is not the number of exposures per roll anymore, it is the time spent selecting the best photo. To get around this bottleneck, we need to automatically assess the quality of images. The scope of this paper is limited to quantifying perceived quality given reference and compared images. However, no-reference quality estimators can be extended with spatiochromatic models to utilize color information in measuring perceived image quality.

The most intuitive characteristic to compare reference and degraded images is fidelity. Mean squared error (MSE) is a commonly used pixel-wise fidelity method, which is calculated by obtaining the difference between reference and distorted images, taking the square root of the difference, and calculating the mean value. MSE is scaled and mapped with a logarithmic function to compute the peak signal-to-noise ratio (PSNR). The authors in \cite{Ponomarenko2011} extend PSNR by adding contrast change and mean shift sensitivity, and quantizing DCT coefficients.

Instead of tracking all the changes in the intensity channel of an image, we can solely focus on sharp changes. Gradient magnitude (GM) is used in image-quality metrics \cite{Zhang2011,Zhang12} to quantify local contrast. The most commonly used operators to calculate GM are Sobel, Prewitt, and Scharr gradient. Sharp changes in intensity are captured by GM but the significance of these changes is not quantified. However, phase congruency (PC) can be used to quantify the perceptual significance of changes. Since GM and PC are complementary, the authors in \cite{Zhang2011} combine them to assess image quality with a method denoted as feature-similarity index (FSIM).

Perceptual significance can also be detected using saliency-based approaches such as spectral residual \cite{Hou2007} and low-level spatiochromatic grouping \cite{Otazu10}. Spectral residual quantifies the difference between reference and distorted images in the frequency domain and rejects information shared by both images. Quantifying the difference mimics the sensitivity of a visual system to unexpected changes and rejecting shared information corresponds to suppression mechanisms in a visual system \cite{Hou2007}. Based on the hypothesis that low-level visual mechanisms are not only responsible for enhancing or suppressing image details but also for detecting salient regions, the authors in \cite{Murray13}  use the spatiochromatic induction model to estimate saliency.

\begin{figure*}[htbp!]
\centering
\includegraphics[width=0.95\linewidth, trim= 0mm 0mm 0mm 0mm]{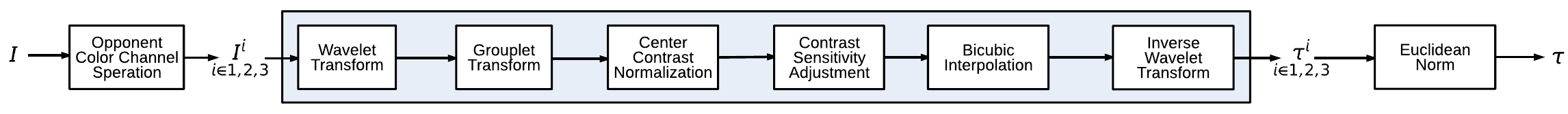}
\vspace{-5.0mm}
\caption{Spatiochromatic grouping pipeline.}
\label{fig:BLeSS_Pipeline_part1}
\vspace{-5.0mm}
\end{figure*}

Saliency can be used to assign local significance to certain regions in images. Spectral residual (SR-SIM) is used to highlight perceptually significant changes in a gradient map to estimate image quality  \cite{Zhang12}. SR-SIM considers sharp changes and  their perceptual significance but it overlooks the color perception in a visual system. A simple way to introduce color perception is pixel-wise fidelity as in FSIMc \cite{Zhang2011} and PerSIM \cite{Temel2015}. Both FSIMc and PerSIM transfer pixel values from the RGB domain to different color spaces where luma and chroma information is separated. FSIMc uses the YIQ space and PerSIM uses the Lab space. Although these quality estimators use color information, they overlook human visual system characteristics. 

To include visual system characteristics in the image-quality assessment, we propose using a biologically-inspired low-level spatiochrmatic-similarity ( \texttt{BLeSS}) method to complement gradient magnitude- and spectral residual-based quality estimators. Main components in the proposed spatiochromatic-similarity method are described in Section \ref{sec:main_bills}, the methodology to use \texttt{BLeSS} as an assistance mechanism is described in Section \ref{sec:main_assistance} and visualization of quality maps is given in Section \ref{sec:vis}. We validate the proposed image-quality assistance method \texttt{BLeSS} in Section \ref{sec:val} and conclude our work in Section \ref{sec:conc}.

\vspace{-3.0mm}

\section{Bio-inspired low-level spatiochromatic model}
\label{sec:main_bills}

The authors in \cite{Otazu08b} introduced a brightness-based low-level induction model (BIWaM) using multi-resolution wavelets. BIWaM was shown to mimic basic perception mechanisms including but not limited to simultaneous contrast, the White effect, grating induction, the Todorovic effect, Mach bands, the Chevreul effect, Adelson-Logvinenko tile effects, and the dungeon illusion. BIWaM also unified the brightness contrast and assimilation effects into a single model. Brightness contrast describes the phenomenon when the brightness of test stimuli shifts away from the surroundings and brightness assimilation is the opposite case when the shift is toward the surroundings. Chromatic induction model (CIWaM), which is an extension of brightness model, mimics the chromatically opponent visual pathways. CIWaM model is based on three main observations, which are partially modeled by the main components (spatial decomposition, surround-contrast model, and contrast sensitivity adjustment) of the proposed spatichromatic similarity method.

Spatial-frequency effect is the first observation, which states that the perception of the central stimuli is influenced by the frequency characteristics of the surround stimuli. The second observation is the spatial-orientation effect, which means that the similarity between the orientation of central and surround stimuli leads to assimilation of the central stimuli whereas difference in the orientation leads to contrast. Finally, the third observation is the surround-contrast effect, which indicates that the contrast of the surrounding stimuli leads to the assimilation of the central stimuli. The authors in \cite{Murray13} extended the chromatic induction model with the low-level spatiochromatic grouping to estimate saliency. In this work, we extend the saliency by induction model as a similarity method to improve the image-quality assessment. In the following subsections, we explain the main blocks in the proposed similarity method.

\vspace{-2.0mm}

\subsection{Introduction to BLeSS }
\label{main0}
\begin{figure}[htbp!]
\vspace{-4.0mm}
\centering
\includegraphics[width=0.95\linewidth, trim= 0mm 0mm 0mm 0mm]{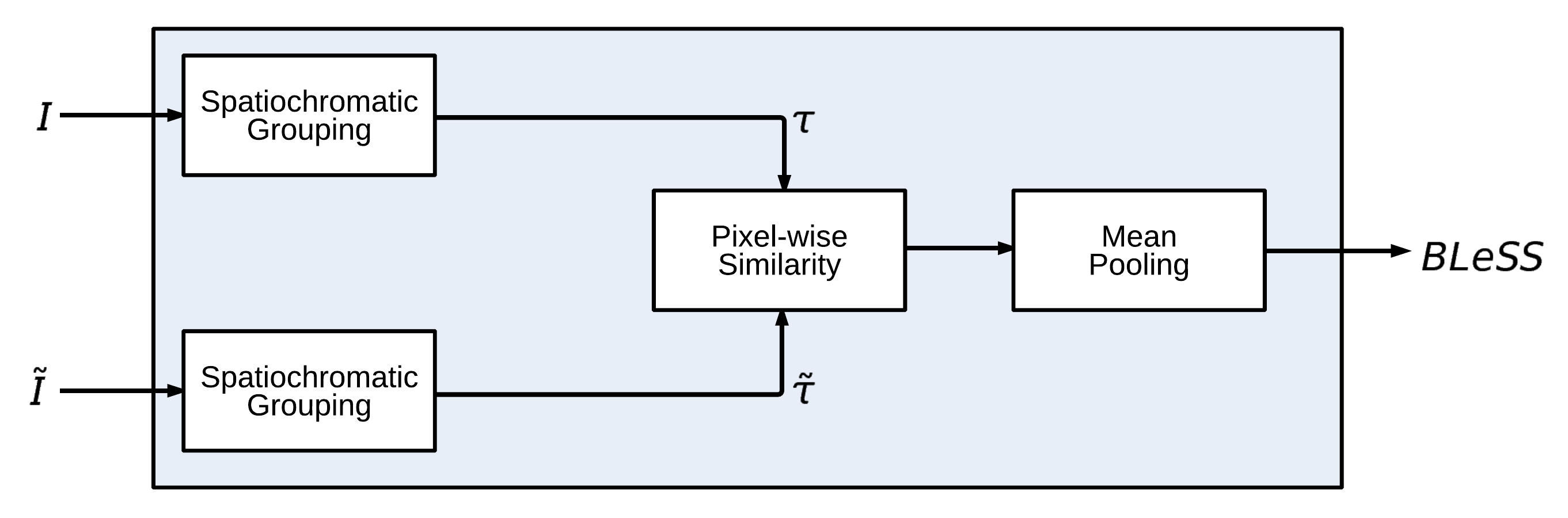}
\vspace{-4.5mm}
\caption{BLeSS pipeline.}
\label{fig:BLeSS_Pipeline_part2}
\vspace{-6.0mm}
\end{figure}

\vspace{-2.0mm}
Images are frequently transformed from the RGB color space to opponent color spaces. For each color channel, we perform following operations. First, wavelet transform is applied to obtain wavelet planes. Then, grouplet transform is applied over wavelet planes. Center-contrast normalization and contrast sensitivity adjustment follow the grouplet transform. Bicubic interpolation is used to obtain the original resolution and inverse wavelet transform is applied to go back to spatial domain. Euclidean norm is used to obtain the spatiochromatic grouping map as summarized in Fig. \ref{fig:BLeSS_Pipeline_part1}. These feature maps are fed to pixel-wise similarity blocks and  mean-pooled to obtain the \texttt{BLeSS} score as given in Fig. \ref{fig:BLeSS_Pipeline_part2}.   

\subsection{Color Domain Transformation}
\label{main1_color_trans}
To partially model chromatically opponent visual pathways, an RGB image is transformed to an opponent color space image, after gamma ($\gamma$) correction, as
\vspace{-2.0mm}
\begin{equation}
\label{eq:main_1_color_trans}
I^1=\frac{R-G}{R+G+B}, I^2=\frac{R+G-2B}{R+G+B},I^3=R+G+B ,
\vspace{-0.5mm}
\end{equation}
where $I^1$, $I^2$ and $I^3$ are  opponent color channels and $R$, $G$, and $B$ are RGB components of a reference image. 
\vspace{-2.0mm}
\subsection{Spatial Decomposition}
\label{main2_wave_trans}
We follow a spatial decomposition approach to obtain frequency and orientation information, which is used in modeling spatial frequency and spatial orientation effects. Spatial decomposition is performed over each color channel map to obtain scale and orientation information as
\vspace{-1.0mm}
\begin{equation}
\label{eq:main2_wave_trans}
\vspace{-1.0mm}
\left\{{w_{s,o}}\right\}_{1\leq s\leq S,o={h,v,d}}, 
\vspace{-0.5mm}
\end{equation}
where $w_{s,o}$ is the wavelet plane at spatial scale $s$ and orientation $o$, horizontal, vertical and diagonal orientations are represented with $h$, $v$ and $d$. Gabor-like basis functions are used in the wavelet transform to mimic receptive fields of neurons in the cortex. We do not directly use Gabor formulations since they do not have a complete inverse transform.
\vspace{-1.0mm}
\subsection{Grouplet tranform}
\label{main3_group_trans}
We use grouplet transform to enhance abstract representations and suppress non-salient features as introduced in \cite{Murray13}. The approximation component (low frequency) of the wavelet plane is initialized as $a_{s,1,o}$ and other scales can be obtained as
\vspace{-2.0mm}
\begin{equation}
\label{eq:main_3_group_trans_1}
a_{s,j+1,o}\left[m,n\right]=\frac{a_{s,j,o}\left[2m-1,n\right]+a_{s,j,o}\left[2m,n\right]}{2},
\end{equation}
where  $m, n$  are pixel indices, $a$ is the approximation (low frequency) component, and $j$ is the scale. The detail component (high frequency) is calculated as
\begin{equation}
\label{eq:main3_group_trans_2}
d_{s,j+1,o}\left[m,n\right]=\frac{a_{s,j,o}\left[2m,n\right]-a_{s,j,o}\left[2m-1,n\right]}{2^j},
\end{equation}
where $d$ is the normalized difference of consecutive approximation components at grouplet scale $j$.

In the Haar transform, approximation and detail coefficients are computed between pairs of consecutive elements. Grouplet transform is also a type of Haar transform but pairs are not necessarily consecutive. Grouplet transform coefficients are paired along the contour that is common to these coefficients. We can consider contour-based pairing as finding points in the direction of maximum regularity. Grouplet plane is obtained by computing detail components $d_{s,j,o}$ for each scale, which can be considered as a sparse representation of complex geometrical structures.

\subsection{Surround-Contrast Model}
\label{main4_sur_contrast}
We perform divisive normalization to partially model center-surround contrast mechanism as
\vspace{-3.0mm}
\begin{equation}
\label{eq:main4_sur_cont}
z_{s,j,o}[m,n]=\frac{(d_{s,j,o}[m,n]^{cen})^2}{(d_{s,j,o}[m,n]^{cen})^2+(d_{s,j,o}[m,n]^{sur})^2},
\end{equation}
where $m$ and $n$ are pixel indices, $d^{cen}$ is the wavelet coefficient of the central region, $d^{sur}$ is the wavelet coefficient of the surround region and  $z$ is the normalized center contrast.

\subsection{Contrast Sensitivity Adjustment}
\label{main5_cont_sens}
We use an extended contrast sensitivity function (ECSF) \cite{Murray13} to model spatial frequency, spatial orientation and surround contrast effects. Normalized coefficients along with spatial frequency and orientation information are used as the input of ECSF as 
\vspace{-4.0mm}
\begin{equation}
\label{eq:main5_cont_sens}
\alpha_{s,j,o}[m,n]=ECSF(z_{s,j,o}[m,n]),
\end{equation}
where $ECSF$ is the extended contrast sensitivity function and $\alpha$ is the contrast-adjusted and divisive-normalized coefficient. ECSF is defined as the summation of two terms. The first term is the multiplication of a normalized coefficient and an approximated psychophysical contrast sensitivity function and the second term is introduced to set the lower bound non-zero. Two different contrast sensitivity functions are obtained from subjective tests, which include brightness and color induction experiments. A more detailed description of the ECSF function is provided by the authors in \cite{Murray13}. 

\vspace{-3.0mm}

\subsection{Interpolation and Inverse Transform}
\label{main6_inter_inverse}
Bicubic interpolation is used to resize each plane ($\alpha_{s,j,o}$)  and these planes are combined to obtain a single plane ($\alpha_{s,o}$). Inverse wavelet is used to transform wavelet coefficients back to spatial domain ($\tau^i$), where $i$ corresponds to the channel index.

\subsection{Pooling and Similarity Computation}
\label{eq:main7_pooling}
Color channels are combined as
\vspace{-3.0mm}

\begin{equation}
\label{main7_pooling}
\tau=\sqrt{\sum_{i=1}^{3}(\tau^i)^2},
\end{equation}
where $\tau$ is the combined map. Same operations are also applied to the distorted image to obtain $\tilde{\tau}$. The similarity between feature maps are calculated with the familiar expression that has been used in most of the pixel-wise and structural similarity metrics as 
\vspace{-4.0mm}

\begin{equation}
\label{eq:sim}
BLeSS=\frac{2\cdot (\tau)\cdot (\tilde{\tau})+C_1}{(\tau)^2+(\tilde{\tau})^2+C_1},
\end{equation}
where $\tau$ is the feature map of the reference image, $\tilde{\tau}$ is the feature map of the distorted image and $C_1$ is a constant added to the denominator to avoid stability issues when the denominator converges to $0.0$ and $C_1$ is also added to the numerator to avoid the bias. Similarity score is $1.0$ when feature maps are same and it converges to $0.0$ as the difference between the compared images increases.

\vspace{-3.0mm}

\section{BL\texorpdfstring{\MakeLowercase{e}}~SS-ASSISTED QUALITY ASSESSMENT}
\label{sec:main_assistance}
\texttt{BLeSS} is used to enhance feature similarity- (FSIM, FSIMc) and spectral residual-based (SR-SIM) quality estimators described in Section \ref{sec:intro}. In all these quality estimators, feature maps are masked with weight maps which are introduced as representations of reliability, saliency, or region of interest. The intuition behind using a weight map is to assign significance to pixels so that when a feature map is pooled into a final quality score, significant pixels would be more influential. However, in practice, these feature maps are usually the same as the ones that are already used in the quality estimators and they are not specifically designed to identify regions of interest. We can formulate the weighting operation as
\vspace{-2.0mm}
\begin{equation}
\label{eq:weight_map}
\frac{\sum_{m=1}^{M} \sum_{N=1}^{N} F[m,n] \cdot W[m,n]}{\sum_{m=1}^{M} \sum_{N=1}^{N} W[m,n]},
\end{equation}
where $m$ and $n$ are pixel indices, $F$ is the feature map and $W$ is the weight map. In the following subsections, we formulate the  \texttt{BLeSS} assisted methods  using the following notation: $GM$  is the gradient magnitude similarity map, $PC$ is the phase congruecny similarity map and $SR$ is the spectral residual similarity map.

\vspace{-2.0mm}

\subsection{BLeSS-FSIM}
\label{sec:fsim}
\texttt{BLeSS} assisted FSIM feature map is defined as
\vspace{-2.0mm}
\begin{equation}
\label{eq:fsim_fm}
F=GM \cdot PC \cdot BLeSS,
\vspace{-2.0mm}
\end{equation}
where $\cdot$ is the pixel-wise multiplication operator. The FSIM weight map is given by
\vspace{-2.0mm}
\begin{equation}
\label{eq:fsim_wm}
W=max(PC\cdot \tau,\tilde{PC\cdot \tilde{\tau}}),
\vspace{-2.0mm}
\end{equation}
where the spatiochromaticly grouped map of the reference image is $\tau$ and the compared map is $\tilde{\tau}$, $max$ operator takes two feature maps as input and outputs a map whose pixels are the max of the pixels in the compared maps.

\subsection{BLeSS-FSIMc}
\label{sec:fsimc}
\texttt{BLeSS} assisted FSIMc feature map is given by
\vspace{-2.0mm}
\begin{equation}
\label{eq:fsimc_fm}
F=GM \cdot PC \cdot real((I \cdot Q)^{C_2}) \cdot BLeSS^{C_3},
\vspace{-2.0mm}
\end{equation}
where $I$ and $Q$ are pixel-wise similarity maps in the $YIQ$ domain, $real$ is the real value operator, $C_2$ and $C_3$ are parameters used to adjust the significance of the color similarity. $C_2$ is set to $0.03$ in the original implementation \cite{FSIM_code} and $C_3$ is set to $0.3$ to assign a higher weight to \texttt{BLeSS} compared to the pixel-wise color similarity. \texttt{BLeSS} assisted FSIMc weight map is defined as
\vspace{-2.0mm}
\begin{equation}
\label{eq:fsmc_wm}
W=max(PC \cdot \tau ,\tilde{PC} \cdot \tilde{\tau}),
\end{equation}
where $max$ is the pixel-wise maximum operator, $PC$ is the phase congruency map, $\tau$ is the spatiochromaticly grouped map of the reference image, $\tilde{PC}$ and $\tilde{\tau}$ correspond to the feature maps of the distorted image.

\subsection{BLeSS-SR-SIM}
\texttt{BLeSS} assisted SR-SIM feature map is defined as
\vspace{-2.0mm}
\label{sec:srsim}
\begin{equation}
\label{eq:srsim}
\vspace{-2.0mm}
F=SR \cdot (GM \cdot BLeSS)^{C_4},
\end{equation}
where $C_4$ is a constant set to $0.5$ in the original implementation \cite{SRSIM_code}. The SR-SIM weight map is given by
\vspace{-1.0mm}
\begin{equation}
\label{eq:srsim_wm}
\vspace{-2.0mm}
W=max(SR\cdot \tau,\tilde{SR}\cdot \tilde{\tau}),
\end{equation}
where $SR$ is the spectral residual map and $\tau$ is the spatiochromatic grouped map of the reference image and the feature maps of the compared images are $\tilde{SR}$  and $\tilde{\tau}$.

\vspace{-2.0mm}

\subsection{Parameter setup}
Similarity maps based on the gradient magnitude, the phase congruency, the spectral residual,  and the low-level spatiochromatic grouping are computed by substituting the feature maps in the similarity formulation in (\ref{eq:sim}). We use original parameters in the publicly available codes for FSIM \cite{FSIM_code}, FSIMc \cite{FSIM_code} and SR-SIM \cite{SRSIM_code}. The parameters in (\ref{eq:sim}) for different feature maps are summarized in Table \ref{param_sim}. The \texttt{BLeSS} parameter is set to the same value with the SR-SIM parameter without any tuning.

\begin{table}[]
\centering
\caption{Parameters in the similarity formulation.}
\label{param_sim}
\begin{tabular}{|l|l|l|}
\hline
\textbf{Similarity}         & \textbf{Metric} & \textbf{Coefficient: C1} \\ \hline
\multirow{2}{*}{$GM$} & FSIM,FSIMc      & 160                      \\ \cline{2-3}
                            & SR-SIM          & 225                      \\ \hline
$PC$                  & FSIM,FSIMc      & 0.85                     \\ \hline
$SR$                  & SR-SIM          & 0.4                      \\ \hline
$\tau$               & BLeSS           & 0.4                      \\ \hline
\end{tabular}
\end{table}
\section{Visualization}
\label{sec:vis}
\vspace{-2.0mm}

\begin{center}
\begin{figure}[htbp!]
\vspace{-2.50mm}

\begin{minipage}[b]{0.45\linewidth}
  \centering
\includegraphics[width=\linewidth, trim= 25mm 85mm 25mm 80mm]{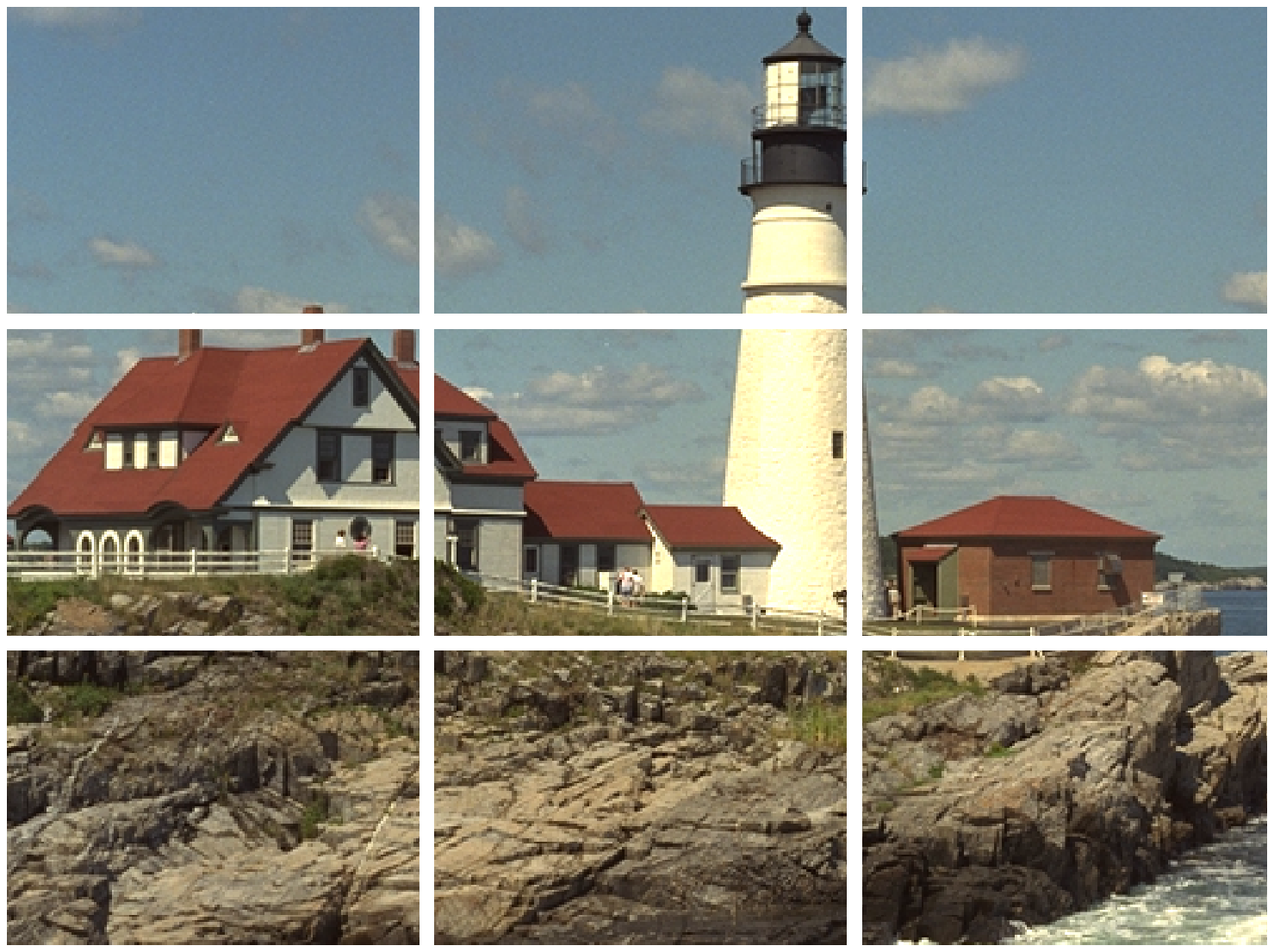}
  \vspace{-0.03cm}
  \centerline{\footnotesize{(a)Pristine Image}}
    \vspace{-0.30cm}
\end{minipage}
  \vspace{0.20cm}
\hfill
\begin{minipage}[b]{0.45\linewidth}
  \centering
\includegraphics[width=\linewidth, trim= 25mm 85mm 25mm 80mm]{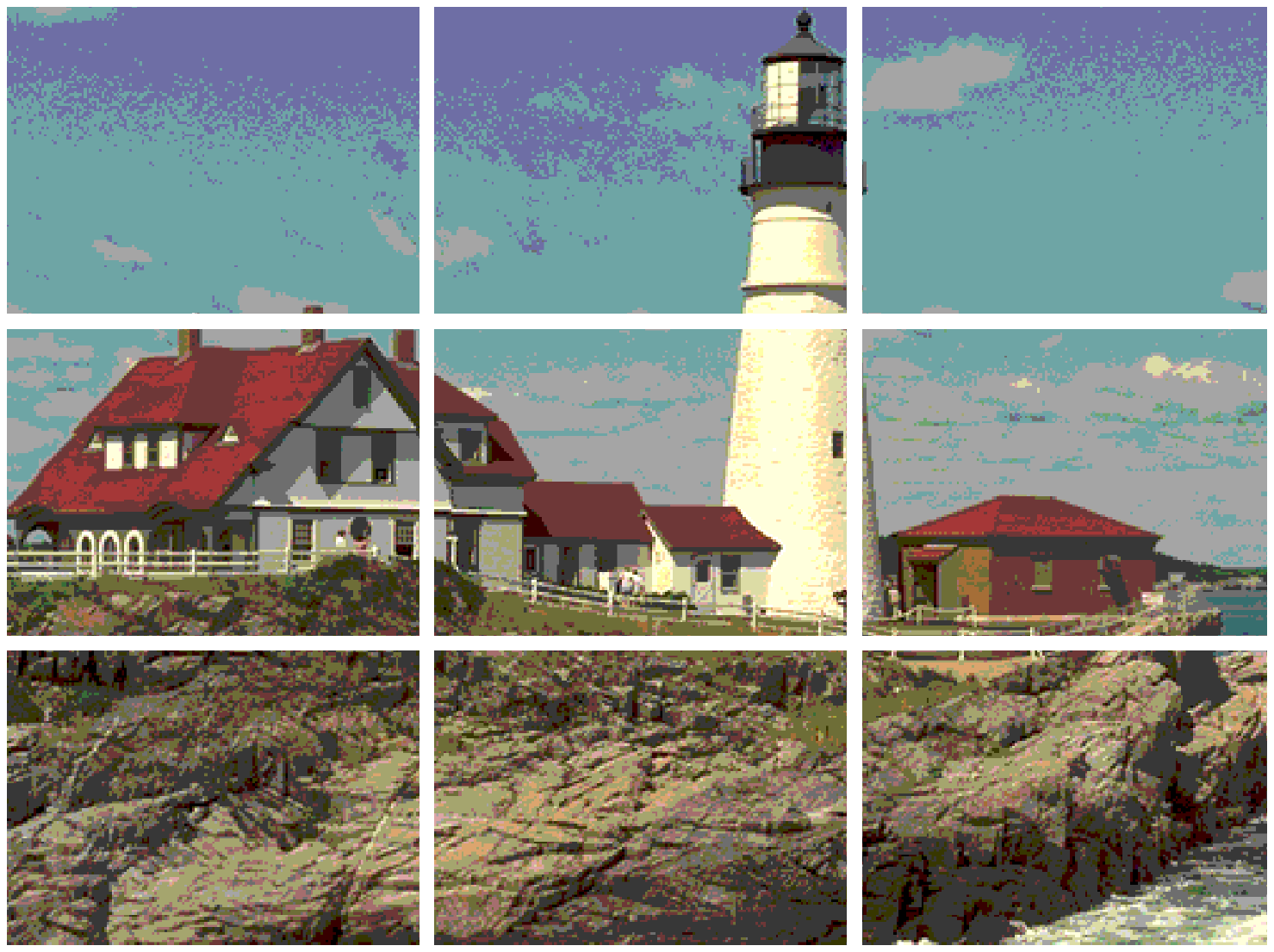}
  \vspace{-0.03cm}
  \centerline{\footnotesize{(b) Distorted Image }}
    \vspace{-0.30cm}
\end{minipage}
  \vspace{0.20cm}
\hfill
\begin{minipage}[b]{0.45\linewidth}
  \centering
\includegraphics[width=\linewidth, trim= 25mm 85mm 25mm 70mm]{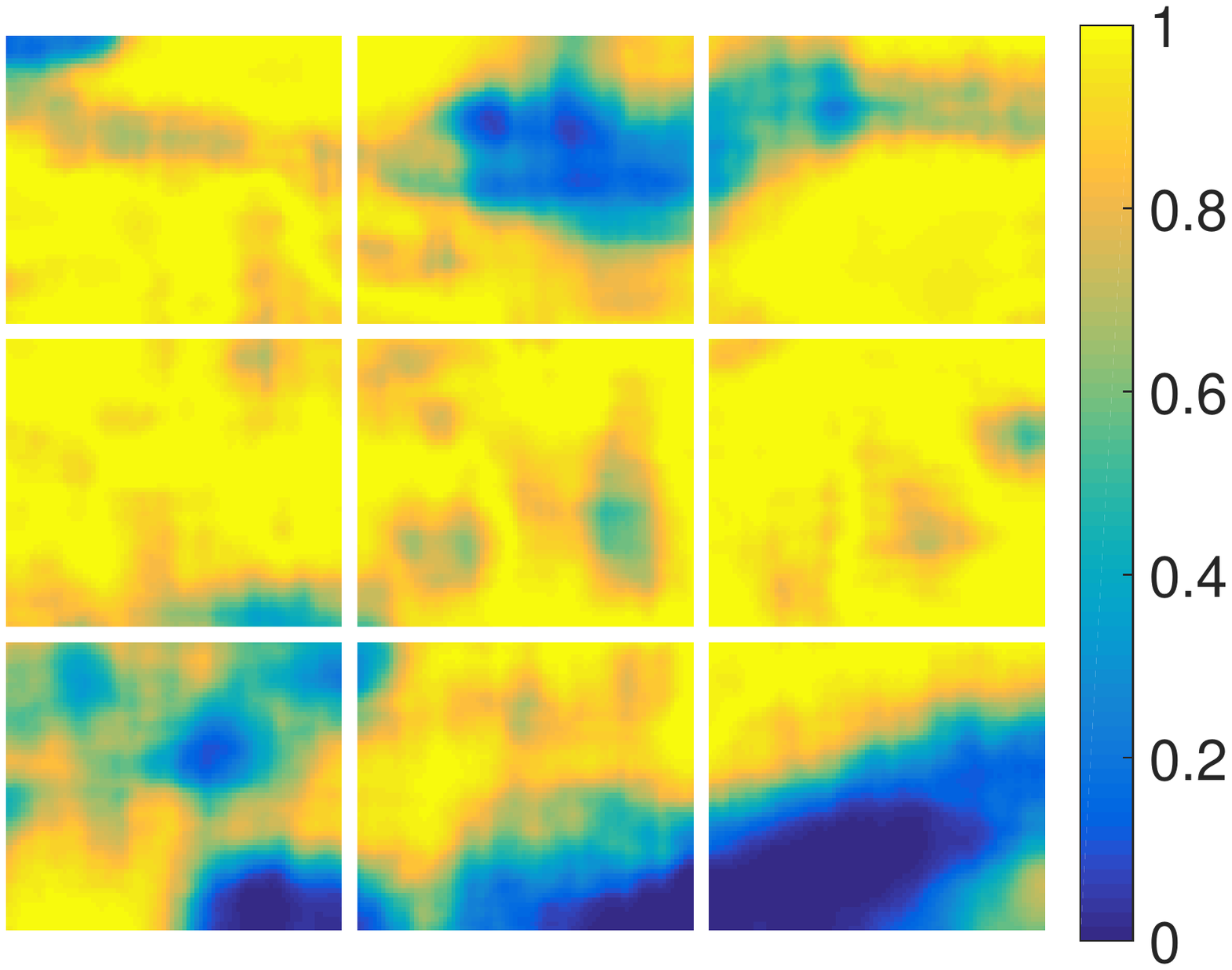}
  \vspace{-0.03 cm}
  \centerline{\footnotesize{(c)  BLeSS Map   } }
      \vspace{-0.30cm}
\end{minipage}
  \vspace{0.20cm}
\hfill
\begin{minipage}[b]{0.45\linewidth}
  \centering
\includegraphics[width=\linewidth, trim= 25mm 85mm 25mm 70mm]{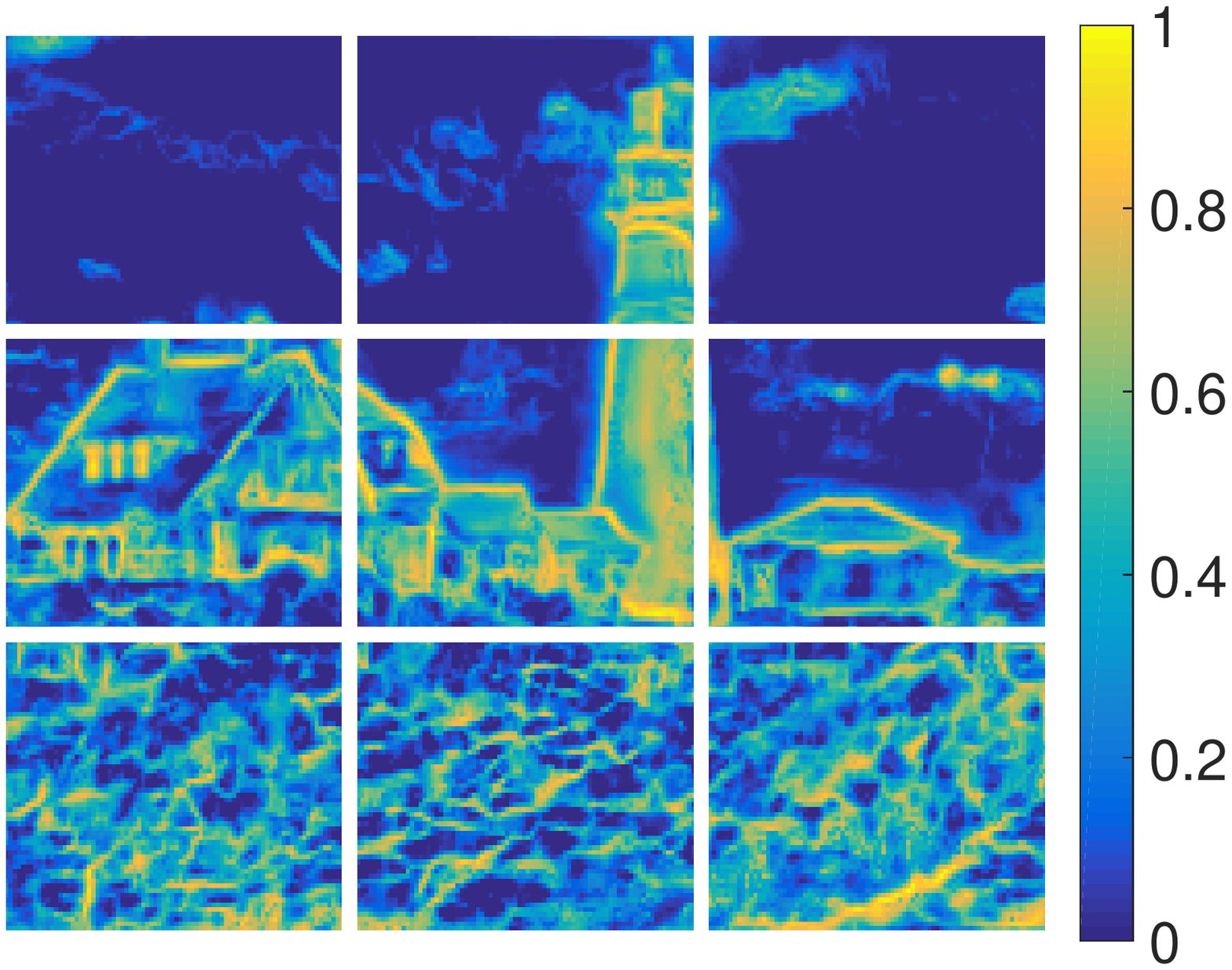}
  \vspace{0.03 cm}
  \centerline{\footnotesize{(d) FSIM Map }}
      \vspace{-0.30cm}
\end{minipage}
  \vspace{0.20cm}
\hfill
\begin{minipage}[b]{0.45\linewidth}
  \centering
\includegraphics[width=\linewidth, trim= 25mm 85mm 25mm 70mm]{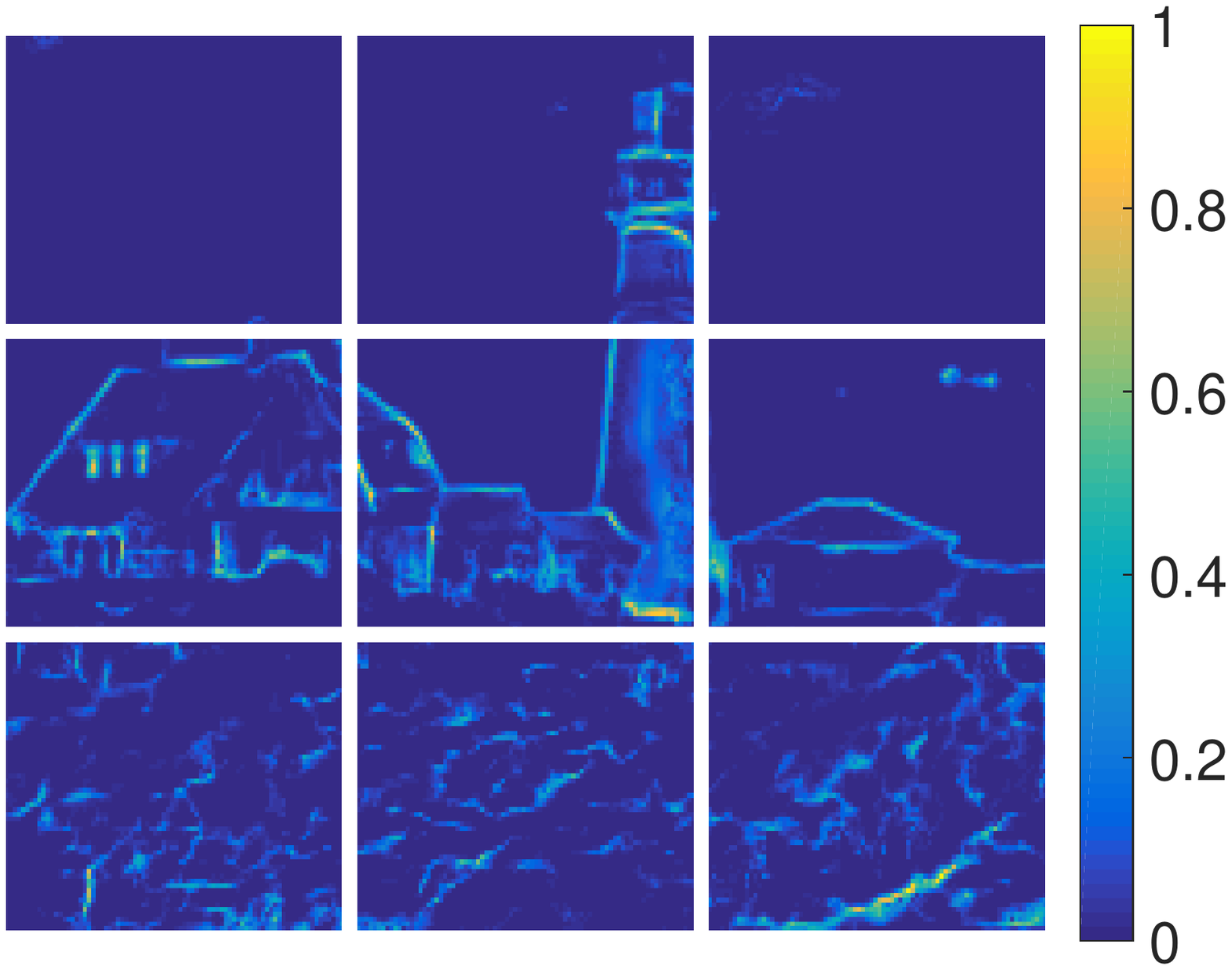}
  \vspace{0.03 cm}
  \centerline{\footnotesize{(e)  FSIMc Map  } }
      \vspace{-0.30cm}
\end{minipage}
  \vspace{0.20cm}
\hfill
\begin{minipage}[b]{0.45\linewidth}
  \centering
\includegraphics[width=\linewidth, trim= 25mm 85mm 25mm 70mm]{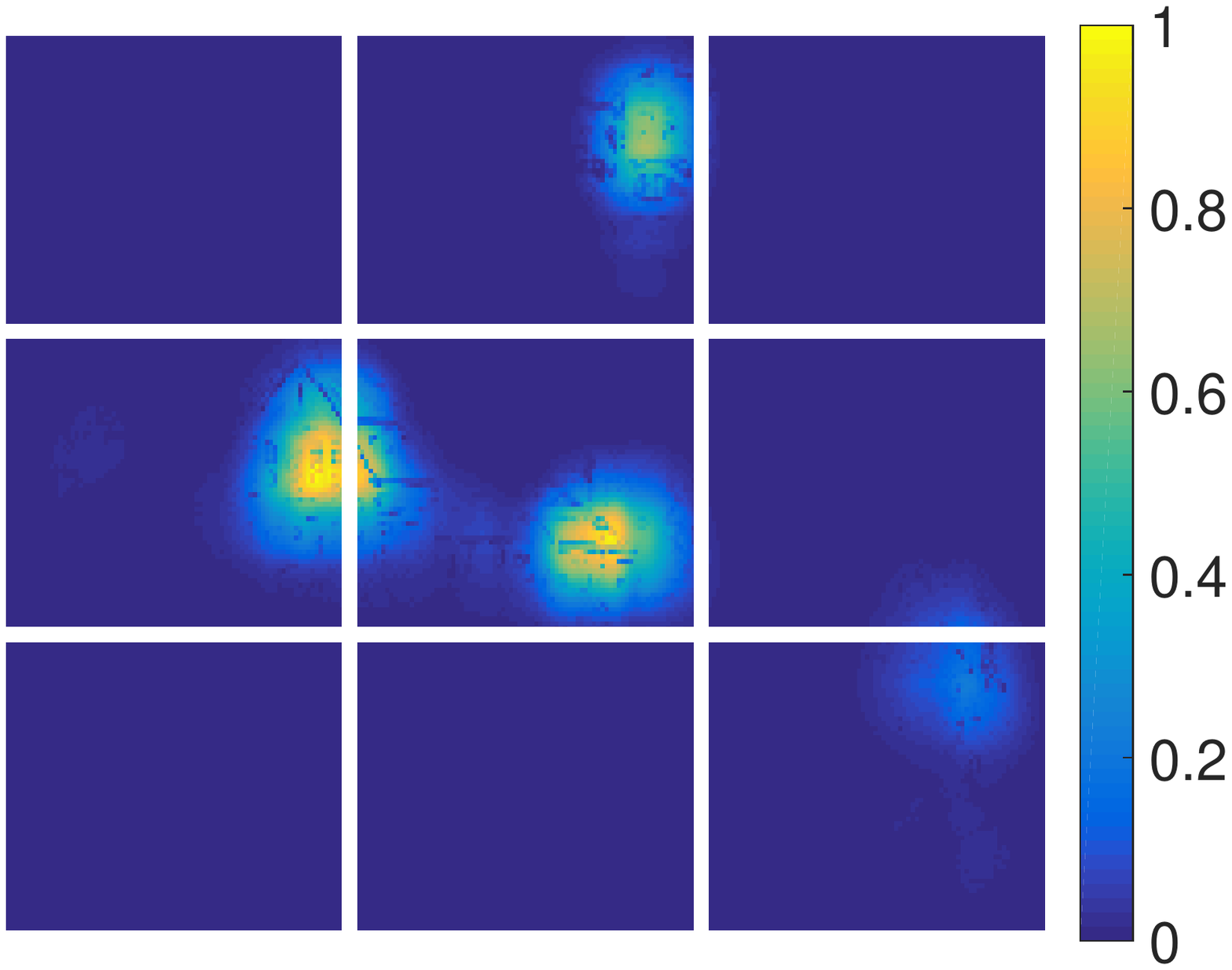}
  \vspace{0.03 cm}
  \centerline{\footnotesize{(f) SR-SIM Map}}
      \vspace{-0.30cm}
\end{minipage}
\vspace{-4.0mm}

\caption{Pristine and distorted images with corresponding quality maps.}\vspace{-.5cm}
\label{fig:visual}
\vspace{-2.0mm}
\end{figure}

\end{center}

In order to visually compare quality maps, we use the \texttt{lighthouse2} image from the TID 2013 database \cite{TID2013} degraded with quantization to illustrate the distortion maps corresponding to each block in the \texttt{BLeSS}, FSIM, FSIMc and SR-SIM maps. We show weighed quality maps which correspond to the numerator of the expression in (\ref{eq:weight_map}).  All images are shown with a grid structure to make the visual comparison easier among images and quality maps. Reference and distorted images are shown in Fig. \ref{fig:visual}(a)-(b). We normalize all feature maps by subtracting the mean, dividing by the maximum and taking the $5th$ power of pixel values to visually highlight the difference between quality maps.

Degradations based on color and structure are significant in the top row, especially in the middle grid. Sharp tone changes and  pixel-wise discontinuities in the sky are easily perceived as well. In the middle row, we can observe degradation over roofs of houses and around windows where we have edges or sharp transitions. However, it is not easy to observe degradations around regions with over exposure such as the wall of the lighthouse. Degradations are less perceivable around the highly textured regions as observed in the  bottom grids where we have the textured rock components.

\texttt{BLeSS} map captures some of the degradations in the sky region, especially the middle grid. In the middle row, only some of the sharp changes are captured and some are overlooked like changes around the big roof. In the bottom row, degradations around the transition between the rock and the ocean are captured but the estimated quality is lower than the sky region which is not accurate. FSIM can capture the degradations that are overlooked by \texttt{BLeSS} but FSIM also captures pixel-wise changes that are not even perceived because of the masking effect around highly textured regions. FSIMc is not oversensitive to all the changes compared to FSIM but it overlooks significant degradations such as the sky region in the top grids. SR-SIM can detect some of the degradations but it is not very sensitive to the level of degradations. SR-SIM identifies four regions as high quality and the rest as low quality in this visual example.

\vspace{-3.00mm}
\section{Validation}
\label{sec:val}
\vspace{-3.00mm}

\subsection{Databases}
\label{sec:val_db}
\texttt{BLeSS} is validated using LIVE \cite{live2006}, Multiply Distorted LIVE (MULTI) \cite{multi2012} and TID 2013 (TID13) \cite{TID2013} databases. All of the distortion types in these databases can be grouped into seven categories. \texttt{Compression} includes Jpeg, Jp2k and lossy compression of noisy images. \texttt{Noise} contains  white noise, adaptive Gaussian noise, additive noise in chroma, impulse noise to simualte acquisition errors, spatially correlated noise to model digital photography error, masked noise and high frequency noise to simulate compression and watermarking error, quantization noise to model registration and gamma correction error, image denoising, multiplicative Gaussian noise, comfort noise and lossy compression of noisy images. \texttt{Communication} includes Rayleigh fast-fading channel model, Jpeg and Jp2k transmission errors. \texttt{Blur} consists of Gaussian blur  and sparse sampling and reconstruction error. \texttt{Color} contains color saturation change, color quantization with dither and chromatic aberrations. \texttt{Global} includes intensity shift to stimulate image acquisition error and contrast change to model image acqustion and gamma correction error. \texttt{Local} consists of non-eccentricity pattern to model image compression and watermarking, local block-wise distrotion of different intensity to simulate inpainting and acquisition errors.
The number of images in each category is summarized in Table \ref{tab_db}.

\begin{table}[htbp!]
\vspace{-2.50mm}
\centering
\caption{The number of distorted images with respect to degradation categories in each database.}
\label{tab_db}
\begin{tabular}{|l|l|l|l|l|}
\hline
                    & {\bf LIVE} & {\bf MULTI} & {\bf TID13} & {\bf Total} \\ \hline
{\bf Comp.}   & 460        & 225         & 375         & 1060        \\ \hline
{\bf Noise}         & 174        & 225         & 1375        & 1774        \\ \hline
{\bf Comm.} & 174        & -           & 250         & 424         \\ \hline
{\bf Blur}          & 174        & 450         & 250         & 874         \\ \hline
{\bf Color}         & -          & -           & 375         & 375         \\ \hline
{\bf Global}        & -          & -           & 250         & 250         \\ \hline
{\bf Local}         & -          & -           & 250         &250          \\ \hline
\end{tabular}
\end{table}

\vspace{-3.0mm}
\subsection{Performance metric}
\label{sec:val_metric}
\vspace{-1.0mm}
The performance of the proposed quality estimator assistance is validated using the Spearman correlation coefficient. We avoid using accuracy- and linearity-based metrics since they rely on the range of estimated scores and the regression function. Spearman's rank correlation coefficient assigns ranks to scores and estiamtes. These ranks are used instead of exact scores. For example, let's assume that we have $N$ images with corresponding mean opinion scores ($y_i$). Based on the rankings, the minimum score should be assigned as $1$, the maximum as $N$ and the others should be in between $1$ and $N$ based on their rankings. This procedure is applied to both subjective scores and estimates. If the relative order of mean opinion scores and objective estimates are same, the correlation should be $1.0$ otherwise it should be lower. The exact formulation of the Spearman correlation coefficient is given by
\vspace{-4.0mm}

\begin{equation}
\label{eq:spearman}
SRCC=1-\frac{6\sum_{i=1}^{N}(X_i-Y_i)^2}{N\cdot (N^2-1)},
\vspace{-2.0mm}
\end{equation}
where $X_i$ is the rank assigned to  score $x_i$ and $Y_i$ is  the rank assigned to mean opinion score $y_i$ that corresponds to image indexed with $i$ and $N$ is the total number of images.

\begin{table}[]
\centering
\caption{Percentage performance changes for BLeSS assisted IQA metrics over various distortion categories.}
\label{tabs_results_categories}
\begin{tabular}{|l|l|l|l|}
\hline
                 & \textbf{SR-SIM} & \textbf{FSIM} & \textbf{FSIMc} \\ \hline

                 \textbf{Comp.}   & {-0.29 (000)}           & \textbf{+0.13 (000)}         & \textbf{+0.28 (000)}          \\ \hline
                 \textbf{Noise}   & {-2.16 (001)}           & {-1.31 (000)}         & {-0.34 (000)}          \\ \hline
                 \textbf{Comm.}   & \textbf{+0.07 (0-0)}           & \textbf{+0.25 (0-0)}         & \textbf{+0.24 (0-0) }          \\ \hline
                 \textbf{Blur}   & {-0.39 (000)}           & \textbf{+0.20 (000)}         & \textbf{+0.40 (000)}          \\ \hline
                 \textbf{Color}   & \textbf{ +183 (--1)}           & \textbf{+185 (--1)}         & \textbf{+13.1 (--1)}          \\ \hline
                 \textbf{Global}   & -1.31 (--0)           & {-4.69 (--0)}         & {-0.16 (--0)}          \\ \hline
                 \textbf{Local}   &-1.85 (--0)           & \textbf{+4.36 (--0)}         & \textbf{+3.12 (--0)}          \\ \hline
\end{tabular}
\vspace{-4.00mm}
\end{table}

\subsection{Results}
\label{sec:val_results}
We analyze the effect of \texttt{BLeSS} by focusing on relative performance changes percentage wise for FSIM, FSIMc and SR-SIM. Distortion category-based relative performance changes are provided in Table \ref{tabs_results_categories}. Results are highlighted if there is an increase in the performance. We calculate performance changes in each database and provide the weighted average. In case of communication distortions, we can see a minor increase for all the quality maps. There are slight increases in the performance for FSIM and FSIMc in compression and blur category and relatively higher increases in local distortion category. In color distortion category, there is more than $100\%$ increase for SR-SIM and FSIM and there is around $10\%$ increase in FSIMc. The increase in FSIMc is less compared to others since color-based similarity is already included in the quality metric but \texttt{BLeSS} still enhances the performance. The overall performance changes in case of \texttt{BLeSS} assistance is given in Table \ref{tabs_results_overall}. The performance of FSIM and FSIMc increase for all databases whereas the performance of SR-SIM increases for the LIVE and the TID13 databases.

We perform statistical tests and analysis to verify that differences in terms of correlation coefficients are not solely random and they are statistically significant. In order to analyze the difference between correlation coefficients, we use statistical significance tests suggested in ITU-T Rec. P.1401. \cite{ITU2012}. In Table \ref{tabs_results_categories} and Table \ref{tabs_results_overall} , we report the statistical significance test results within parentheses next to the percentage change. In these test results, a $0$ means that the change is not statistically significant whereas a $1$ corresponds to a statistically significant change. In Table \ref{tabs_results_categories}, we provide the statistical significance for each distortion type and database. The first index corresponds to the LIVE, the second index is for the MULTI and the third is for the TID13. If a specific database does not include a distortion type, there is a hyphen. The decrease in the performance of SR-SIM in the noise category of TID13 database is low. However, it is still statistically significant since the decrease is for 1,375 images. Moreover, the performance enhancement in color category is significant for all of the quality estimators. As summarized in Table \ref{tabs_results_overall}, in full databases, the increases in FSIMc are not statistically significant whereas increases in SR-SIM and FSIM are statistically significant in the TID2013 database.

\begin{table}[htbp]
\centering
\vspace{-3.00mm}
\caption{Percentage performance changes for BLeSS assisted IQA methods over full databases.}
\label{tabs_results_overall}
\begin{tabular}{|l|l|l|l|}
\hline
                 & \textbf{SR-SIM} & \textbf{FSIM} & \textbf{FSIMc} \\ \hline
\textbf{LIVE}    &\textbf{+0.13 (0)}           & \textbf{+0.17 (0)}         & \textbf{ +0.06 (0)}           \\ \hline
\textbf{MULTI}   & { -0.33 (0)}            & \textbf{+0.62 (0)}         & \textbf{+0.79 (0)}          \\ \hline
\textbf{TID13}   & \textbf{+3.79 (1)}           & \textbf{+4.77 (1)}         & \textbf{+1.03 (0)}          \\ \hline
\end{tabular}
\vspace{-3.00mm}
\end{table}

\section{CONCLUSION}
\label{sec:conc}
\vspace{-3.5mm}
We proposed an assistance similarity method based on a bio-inspired low-level spatiochromatic grouping model to partially mimic the spatial frequency, spatial orientation and surround contrast effects in the perceptual quality assessment. The proposed assistance similarity method \texttt{BLeSS} is used to enhance image-quality methods that originally overlook or oversimplify the perception of color in the visual quality assessment. The results in the LIVE, the Multiply Distorted LIVE and the TID 2013 databases show that \texttt{BLeSS} increases the quality assessment performance for feature similarity metrics in all the databases and in the LIVE and the TID2013 databases for spectral residual-based metric. In terms of statistical significance, changes in most of the distortion categories are not significant except color and noise. Significant changes in the color category lead to more than $100\%$ enhancement in terms of the Spearman correlation for FSIM and SR-SIM.


\end{document}